# TEMPERATURE VERSUS DENSITY EFFECTS IN GLASSFORMING LIQUIDS AND POLYMERS: A SCALING HYPOTHESIS AND ITS CONSEQUENCES

Christiane Alba-Simionesco[1] and Gilles Tarjus[2]

*(1) Laboratoire de Chimie Physique, UMR 8000, bât 349 Université Paris Sud XI, 91405 Orsay cedex, France*

*(2) Laboratoire de Physique Théorique de la Matière Condensée*
*Université Pierre et Marie Curie, 4 Place Jussieu, 75252 Paris Cedex 05, France*

ABSTRACT

We discuss the validity and the outcome of a scaling hypothesis proposed by us some years ago, according to which the influence of the density on the slowing down of flow and relaxation in glassforming liquids and polymers is described trough a single effective interaction energy $E_{\infty}(\rho)$. We stress the formal consequences and the physical meaning of the scaling.

I- INTRODUCTION

The dramatic slowing down of flow and relaxation that leads to glass formation in liquids and polymers as temperature decreases is still a much debated topic among chemists and physicists. However, a distressing aspect of the problem is that there is not even agreement on a zeroth order description of the phenomenon.

For instance, the question of whether slowing down results (1) from congestion of the molecules and drainage of free volume as density increases with decreasing temperature under usual isobaric conditions, or (2) from the decrease of thermal energy required to surmount higher and higher activation barriers as temperature is lowered (or from both effects) is not settled. At one end of the spectrum, free-volume theories and hard-sphere models stress the density effect whereas at the other end, approaches in terms of activated processes emphasize the temperature factor. Repulsive colloidal systems most certainly fall into the category described by the former theories and covalently bonded network-forming glassformers most certainly in that described by the latter; but what about most glassforming molecular liquids and polymer melts? It is reasonable to expect that both density and temperature play a role, and this can be quantitatively assessed by studying ratios characterizing in a model-free manner the relative contributions of density and temperature to the experimentally observed slowing down of relaxation at constant pressure. Several, ratios more or less equivalent, have been proposed[1,2]. From the available data, it appears that temperature on the whole is more important than density (see Fig.1 in ref [3,4]), but except in a few cases such as glycerol[5], the dominance is far from overwhelming (and the opposite effect has also been observed in a few instances[6]).

An interesting piece of information, however, is provided by the evolution with increasing relation time (or viscosity) of the relative importance of density and temperature in the slowing down. Data are rather scarce, but all show a decreasing influence of the density as one approaches the glass transition (whatever the thermodynamic path followed)[2,7]. This observation, which is clearly at odds with a free-volume picture[8], suggests that the proposed ratios somewhat keep the memory, even at low temperature close to the glass transition, of the high-temperature liquid phase in which density and temperature are more likely to play a comparable role. This led us a few years ago [9] to propose a simplifying scaling hypothesis in which the influence of density of the $\alpha$-relaxation time $\tau_\alpha$ or the viscosity $\eta$ is simply described via a single parameter, an effective density-dependent interaction energy $E_\infty(\rho)$ that is characteristic of the high-temperature liquid regime[9,3].

In the following, we first recall the scaling procedure and its application to experimental and numerical data on glassforming liquids and polymers. We next consider what use can be made of the existence of such a scaling and derive a number of relations that automatically follow for the scaling form. Finally, we discuss the physical content of the scaling and of the density-dependent energy parameter, including some recently proposed power-law dependence.

II- SCALING PROCEDURE

The proposed scaling hypothesis leads to write the α-relaxation time or the viscosity in the following form[9,3],

$$\log\left(\tau_\alpha(\rho,T) \text{ or } \eta(\rho,T)\right) = \Psi\left[\frac{E_\infty(\rho)}{T}\right], \quad (1)$$

where $\tau_\alpha$ is expressed, say, in seconds ($\eta$ in Poise), $E_\infty(\rho)$ is expressed in units for which $k_B \equiv 1$, and $\Psi(X)$ is a species-specific scaling function. A formula similar to Eq. (1) can be used for the effective activation energy $E(\rho,T)$ obtained by describing the T-dependence of $\tau_\alpha$ or $\eta$ in an activated form, $\tau_\alpha(\rho,T)/\tau_\alpha^\infty = \exp[E(T,\rho)/T]$, with negligible ρ and T dependences of the prefactor $\tau_{\alpha,}^\infty$:

$$E(\rho,T)/E_\infty(\rho) = \Phi\left[\frac{E_\infty(\rho)}{T}\right], \quad (2)$$

where $\Phi(X)$ is a scaling function simply related to $\Psi(X)$, leading to

$$\Psi(X) = \ln(\tau_\alpha^\infty) + X\Phi(X) \quad (3)$$

An operational way to determine $E_\infty(\rho)$ uses the fact that for most glassforming liquids above some crossover temperature $T_*(\rho)$ (called e.g. "onset temperature" in ref.[10]), an approximate Arrhenius T-dependence of the relaxation time and viscosity is observed (whereas super-Arrhenius behavior occurs below $T_*(\rho)$). $E_\infty(\rho)$ then naturally identifies to the effective activation energy of the Arrhenius dependence : $\Phi(X)$ goes to 1 below some value of the argument. One should stress at this point that no satisfactory theoretical explanation of the Arrhenius activated behavior has been proposed for the high-T liquid phase; therefore, we take the Arrhenius form as a convenient, but effective description, which physically makes sense only if the empirical $E_\infty(\rho)$ is several times the thermal energy T.

The above procedure for determining $E_\infty(\rho)$ is however inoperative in the case of most polymer melts for which high-T data in the Arrhenius regime are not available. One must then resort to given functional forms for describing the ρ-dependence of $E_\infty(\rho)$. Simple one-parameter formulas are provided by a power law, $E_\infty(\rho) \propto \rho^x$ [13,3,11,12], and a linear behavior, $E_\infty(\rho) \propto \rho - \rho^*$ [3,11] (where in both cases a ρ-independent amplitude factor is left undetermined and is implicitly included in the definition of the scaling functions Ψ and Φ). Better descriptions could be achieved by adding more parameters; for instance, a quadratic term could be added to the linear form in order to describe, if necessary, a curvature in the shape of $E_\infty(\rho)$.

Data collapse on mastercurves, as predicted by Eq. (1), is illustrated in Fig.1a,b,c for different systems[9,3,11]: the binary Lennard-Jones model studied by simulation, for which $E_\infty(\rho)$ had been previously determined[10], the organic glassformer o-terphenyl and a polymer, polyvinylmethylether (PVME). We have also obtained a good data collapse for other liquid models, molecular liquids and polymers. Analysis of the work done by other groups[12,13] confirms the validity of the scaling described by Eqs (1,2).

III- CONSEQUENCES OF THE SCALING

One of course hopes that more work will be undertaken to increase the (ρ,T) data base on a variety of glassformers (with a variety of experimental probes) and to provide additional checks of the data collapse on mastercurves, but taking now the scaling hypothesis for reasonably well established one may explore its consequences. At a theoretical level, the most important outcome is that density can be properly scaled out and the viscous slowing down then described as a thermally activated process. In particular, the isochoric fragility, *i.e*, the measure of the degree of super-Arrhenius behavior at constant density, is independent of the density and can thus be taken as an intrinsic property of a given glassformer (which has led us to suggest a modification of the celebrated Angell plot[11] as illustrated in Fig. 2). This can be seen by considering for instance the isochoric steepness index $m_\rho(\rho,\tau)$ at a given value τ of the relaxation time,

$$m_\rho(\rho,\tau) = \frac{\partial \log(\tau_\alpha(\rho,T))}{\partial(T_\tau/T)}\bigg|_\rho \quad (T = T_\tau), \tag{4}$$

where $T_\tau(\rho)$ is the temperature at which $\tau_\alpha(\rho, T_\tau) = \tau$ (usually, τ~$10^2$sec and $T_\tau$ is simply Tg). As a consequence of Eq.(1), can be rewritten as

$$m_\rho(\rho,\tau) = X_\tau \Psi'(X_\tau), \tag{5}$$

where $\Psi'(X)$ is the first derivative of the scaling function $\Psi$; $X_\tau$ is equal to $E_\infty(\rho)/T_\tau(\rho)$ and satisfies, by definition, $\Psi(X_\tau) = \log \tau$, *i.e.*, is independent of $\rho$. As a result $m_\rho(\rho,\tau)$ is independent of density, which completes the proof.

The mere existence of the scaling formula, Eq.(1) (or Eq.(2)), also generates a number of inter-relations between quantities that would otherwise be independent. For instance, the ratio of the isobaric and isochronic coefficients of expansivity, $\alpha_P/|\alpha_\tau|$, that has been introduced to characterize the relative importance of $\rho$ over that of T in the slowing down at a given pressure (and is simply related to another ratio, $H_P/E_V = [\partial \ln(\tau_\alpha)/\partial(1/T)]_P / [\partial \ln(\tau_\alpha)/\partial(1/T)]_V$, introduced earlier by Williams et al and by Naoki et al[1], can be expressed as

$$\alpha_P / |\alpha_\tau| = \alpha_P T_\tau \frac{d\ln(E_\infty(\rho))}{d\ln(\rho)} \tag{6}$$

where $T_\tau(\rho)$ is defined as above and $d\ln(E_\infty(\rho))/d\ln(\rho)$ reduces to a constant x in the case of a power-law description of the activation energy, $E_\infty(\rho) \propto \rho^x$. Fragility at constant pressure can be characterized by the isobaric steepness index

$$m_P(P,\tau) = \frac{\partial \ln(\tau_\alpha(P,T))}{\partial(T_\tau/T)}\Big|_P \quad (T = T_\tau) \tag{7}$$

where $T_\tau(P)$ is the temperature at which $\tau_\alpha(P,T_\tau) = \tau$. By using Eq.(7), it is easy to show that the isochoric and isobaric steepness indices are related through

$$m_P(P,\tau)/m_\rho(\tau) = 1 + \alpha_P/|\alpha_\tau| = 1 + \alpha_P(P,\tau)T_\tau(P)\frac{d\ln(E_\infty(\rho))}{d\ln(\rho)}. \tag{8}$$

The above equation shows immediately that even if the isochoric fragility is independent of density, the isobaric one depends on pressure[14] and that the isobaric fragility is always larger than the isochoric one, which is indeed experimentally observed. As seen in Fig.2, this latter property leads to a spectrum of fragilities that is reduced in the isochoric case, compared to the isobaric one. We display in Fig.3 for the glassforming polymer PVME the evolution of the ratio $\alpha_P/|\alpha_\tau|$ and of the product $\alpha_P T_g$ at the dielectric glass transition ($T_g \equiv T_\tau$, $\tau_\alpha = 100$ sec) as a function of the pressure $P_g(T)$ or temperature $T_g(P)$. Figure 3 illustrates the following points: (i) $\alpha_P/|\alpha_\tau|$ and $\alpha_P T_g$ are roughly proportional (with a coefficient $x \sim 2.8 \pm 0.2$, close to the value found in Fig.1) and (ii) $\alpha_P/|\alpha_\tau|$ and $\alpha_P T_g$ are not constant but vary by about 30% over the available pressure range. These observations are valid for dielectric and

calorimetric (not shown here) data, however the coefficient *x* is different for the two sets. This discrepancy between different experimental techniques has also been observed for glycerol when comparing dielectric and viscosity measurements[9,5]. It does not contradict the scaling *per se*, but casts doubt on the universality of the temperature dependence of the α relaxation, at least in some materials and for some probes.

Additional "fragilities" have been introduced to characterize the isothermal behavior as a function of density or pressure[15]. Although we are reluctant to extend the concept of "fragility" to such cases, isothermal steepness indices can be defined along those lines (we use different notations to emphasize that one changes both the varying and the constant variables):

$$M_T^{(\rho)}(T,\tau) = \frac{\partial \ln(\tau_\alpha(T,\rho))}{\partial(\rho_\tau/\rho)}\Big|_T (\rho = \rho_\tau), \qquad (9)$$

and a similar definition for $M_T^{(P)}(T,P)$; $\rho_\tau(T)$ and $P_\tau(T)$ are, respectively, the density and the pressure at which $\tau_\alpha = \tau$. By using the scaling formula, Eq. (1), it is easy to derive that

$$M_T^{(\rho)}(T,\tau) = m_\rho(\tau)\frac{d\ln(E_\infty(\rho))}{d\ln(\rho)}, \qquad (10a)$$

$$M_T^{(P)}(T,\tau) = m_\rho(\tau)P_\tau(T)\kappa_T(T,\tau)\frac{d\ln(E_\infty(\rho))}{d\ln(\rho)}, \qquad (10b)$$

where $\kappa_T$ is the isothermal compressibility. Again, using a simple power-law form for $E_\infty(\rho)$ amounts to take $d\ln(E_\infty(\rho))/d\ln(\rho)$ as a species-specific constant (*x*). This little exercise shows that, with one intrinsic fragility $m_\rho(\tau)$, the ρ-dependence of the energy scale $E_\infty(\rho)$, and thermodynamic input, one can retrieve all the various ratios and indices introduced in the literature to characterize the slowing down of relaxation and flow in glassforming liquids and polymers.

Exploring further the consequences of the scaling behaviour, one may obtain constraints on theoretically motivated or simply empirical formulas describing the evolution with temperature of the α-relaxation time or the viscosity. For instance, if a description of $\tau_\alpha(\rho, T)$ introduces a crossover between different functional forms (*e.g.*, a crossover between Arrhenius and super-Arrhenius behavior as in the frustration-limited domain theory[16]), the crossover point for the various isochores, say $T^\#(\rho)$[3,17], must correspond to an isochrone, $\tau_\alpha(T^\#(\rho))$= constant, and its ρ-dependence is simply that of $E_\infty(\rho)$. Indeed, for the change of functional form to be compatible with the scaling, the scaling function Ψ in Eq.(1) (or Φ in Eq.(2)) must change at a given $X^\#$ which, by definition, is independent of ρ and is equal to $E_\infty(\rho)/T^\#(\rho)$ ; as a result, $T^\#(\rho) = E_\infty(\rho)/X^\#$ and the associated relaxation time (or viscosity) is equal to $\Psi = (X^\#)$. Another example is provided by the Adam-Gibbs approach, in which

$$\tau_\alpha(\rho,T) = \tau_0 \exp(A(\rho)/TS_C(\rho,T)) \quad . \tag{11}$$

If the $\rho$-dependence of the prefactor $\tau_0$ is negligible and if $A(\rho)$, which (up to a factor ln(2) in the original version or a similar constant) is an elementary activation energy, scales as $E_\infty(\rho)$, the configurational entropy $S_c(\rho,T)$ must be a function of the scaling variable $X = E_\infty(\rho)/T$ only, *i.e.*,

$$S_c(\rho,T) = \Theta\left[\frac{E_\infty(\rho)}{T}\right] \quad . \tag{12}$$

Finally, the existence of a scaling property may be useful for practical purposes such as: organizing the data collected along various thermodynamic paths in a rational way, predicting by extrapolation from a restricted data base and an estimated scaling function the behavior in a wide range of thermodynamic parameters (provided an accurate equation of state is available). Applied to the case of glassforming molecular liquids that easily crystallize under pressure, this latter procedure could allow, by combining pressure studies at high temperature, which give access to the effective Arrhenius activation energy $E_\infty(\rho)$, and a single isobaric measurement at atmospheric pressure over the whole temperature range down to the glass transition, to determine the glass transition line and more generally all isochronic lines in the whole density-temperature (or pressure-temperature) plane.

IV- PHYSICAL INTERPRETATION OF THE SCALING

We have already stressed that the proposed scaling hypothesis reduces the effect of the density on the dynamics to the variation of a single parameter $E_\infty(\rho)$ that provides the bare scale for activation barriers. The fact that the hypothesis is indeed verified (at least to a good approximation) in real and model glassformers is by no means trivial, because it implies that important aspects of the slowing down of flow and relaxation, such as fragility, are independent of density. Interestingly, one can build an elementary model for which the validity of the scaling can be proven (as suggested in ref 20). The model has been introduced to study the energy landscape of a simple atomic liquid[18]. The potential energy function $U_N$ of the whole system is taken as the sum of repulsive soft-sphere interactions and of a homogeneous, mean-field attractive background. Specifically,

$$U_N(r^{3N}, \rho) = \sum_{i<j=1}^{N} \varepsilon(\sigma/r_{ij})^n - Na\rho \,, \qquad (13)$$

where N is the number of atoms and $a>0$. In the absence of the attractive term, the self-similar nature of the repulsive power-law potential leads to the remarkable property that a uniform rescaling of the atomic coordinates does not modify the topology of the potential energy landscape. This implies the well known result[19] that both the thermodynamic and the (long-time) dynamic properties of soft-sphere model depend on density and temperature only via a single control parameter $\Gamma = \rho T^{-3/n}$. Stated slightly differently, this means that a change of density only changes the bare energy scale, which varies as $\rho^{n/3}$. It is easy to see that adding a uniform attractive background does not alter the dynamical properties (which depend on the topology of the energy landscape and on activation barriers, *i.e.*, on energy differences), but changes the thermodynamic behavior, albeit in a trivial way as illustrated by the equation of state which becomes

$$\frac{P}{\rho T} = F_{repulsive}(\rho T^{-3/n}) - \frac{a\rho}{T} \,, \qquad (14)$$

where F is the scaling function for the purely repulsive soft-sphere model. In this example, Eq.(1) applies with $E_\infty(\rho) \propto \rho^{n/3}$.

Although the soft-sphere model, possibly supplemented by a mean-field attractive background, has been used as a rationale for a universal application of the scaling formula Eq.(1) with a power-law density dependence[20], we think that the model is too much of a caricature for real molecular and polymeric glassformers to guarantee the physical meaning of the power-law exponents (called $x$ here and in ref 10, $\gamma$ in ref 11, and $n$ in ref 12 ). Arguments supporting our view can be summarized as follows (additional arguments are given in ref[21] ):

(1) the power law scaling is valid for spherically symmetric particles, which is obviously not the case for real molecular glassforming liquids; decomposing the intermolecular interactions in site-site atomic-like potentials will not help because the intramolecular constraints will destroy the simple power-law scaling. The presence of directional interactions such as hydrogen bonding makes matter even worse.

(2) In the case of polymers, it is clear that the torsion energy barriers hindering the motion of the monomers introduce an intrinsic energy scale that cannot be accounted for by a power-law density dependence.

(3) Some low values of the power law exponent obtained from scaling collapses (0.13 in the case of sorbitol[12]) are meaningless in an interpretation in terms of soft spheres because it is not compatible with a proper thermodynamic limit (the repulsive energy when integrated over the whole volume diverges).

(4) In the range of densities experimentally accessible, the variation of $E_\infty(\rho)$, when it has been possible to extract it from the scaling procedure[3], appears monotonic and rather featureless. As a result several simple functional forms can reproduce its behavior and lead to good data collapse. We have already mentioned the power-law $\rho^x$ and the affine function $(\rho-\rho^*)$ which in many cases provide a comparable description[11]. The affine function misses possible curvature effects[22]. Those could be introduced at the expense of additional adjustable parameters. However, unless a much wider domain of densities is probed and a strong variation of the energy scale $E_\infty(\rho)$ observed, there are no compelling arguments for choosing one specific functional form and no particular physical meaning to be associated with one form or another.

As an additional check of the applicability of the repulsive soft-sphere model with mean-field attractive background, we have studied the thermodynamics of several systems. We have compared the best available equation of state to the prediction of the model, Eq.(14). More specifically, we have used $a$ and $x$ (*i.e.*, $x=n/3$) as adjustable parameters to obtain the best data collapse of $(P/\rho T + a\rho/T)$ as a function of $\rho T^{-1/x}$. The resulting plots are illustrated in Fig.4 for the case of o-terphenyl: $x=4$ (found in the scaling fit the viscosity data[23]) is acceptable, but does not lead to the best collapse. Again, no compelling argument is found to justify the physical interpretation of the power law $\rho^x$ with $x=4$ (in the present case, no compelling argument is found against it either).

## V- CONCLUSION

We have discussed the validity and the consequences of the scaling hypothesis proposed some years ago[9], according to which the influence of the density on the slowing down of flow and relaxation in glassforming liquids and polymers can be described through a single parameter $E_\infty(\rho)$. This latter represents a bare activation-energy scale characteristic of the high-temperature behavior and possibly reflecting the minimal cooperativity of the molecules with their first neighbors[24]. We have stressed that the mere existence of a unique scaling variable $E_\infty(\rho)/T$ implies additional relations between a variety of dynamic quantities. Finally, we have discussed what we think to be the physical meaning of the scaling, namely the existence of a single ρ-dependent interaction energy scale already set in the high-temperature liquid (when accessible), and cautioned that a power-law description of the density dependence of $E_\infty(\rho)$, convenient as it may be, may not carry much physical content.

ACKNOWLEDGMENTS

We thank Professor Böhmer and Professor Alegria for providing us, respectively, with the viscosity data under pressure for ortho-terphenyl and the dielectric data of PVME. LPTMC and LCP are UMR 7600 and 8000, respectively, at the CNRS.

FIGURE CAPTIONS

Figure 1: Collapse of isochoric data for the α-relaxation time or the viscosity onto a mastercurve, function of the scaling variable $E_\infty(\rho)/T$ : (a) simulation results for the α-relaxation time of the binary Lennard-Jones model[10] ($\tau_{ref}$ is the usual Lennard-Jones unit), (b) viscosity measurements for o-terphenyl[23] ($\eta_{ref}$ is 1 MPa.sec), (c) dielectric relaxation time for poly(vinylmethylether) (PVME)[11] ($\tau_{ref}$ is 1sec). The original data shown in the insets have been collected along isochores for the binary Lennard-Jones, and isotherms from 303K to 423K for o-terphenyl and from 247K to 358K for PVME. The straight lines in (a) and (b) represent the best Arrhenius fit with a slope $E_\infty(\rho)$; in (c) the collapse is done with $E_\infty(\rho) \propto \rho^x$ and $x$=2.7, and a comparable collapse is obtained with $E_\infty(\rho) \propto (\rho-\rho*)$ and $\rho*$=0.62.

Figure 2: Modified Angell plot : $\log(\tau_\alpha(\rho,T))$ versus $X/Xg$, where $X = E_\infty(\rho)/T$ is the scaling variable introduced in the text and $Xg$ its value when $\tau_\alpha$ reaches a characteristic "glass transition" value, say $\tau_{\alpha g}$ = 100*sec* for dielectric relaxation data or $\eta_g$ = $10^{13}$ mPa.s, for several systems; the celebrated Angell plot introduced in this conference 20 years ago is shown in inset, where $\tau_\alpha$ or $\eta$ at atmospheric pressure is plotted versus the inverse scaled temperature Tg/T.

Figure 3: Pressure dependence of the ratio of isobaric to isochronic expansivities $\alpha_P/|\alpha_\tau|$ (full line and left scale) and of the product $\alpha_P Tg(P)$ of PVME (dashed line and right scale) along an isochrone at the dielectric glass transition temperature Tg≡$T_\tau$ (100sec). The data are plotted as function of both $T_g(P)$ and $P_g(T)$.

Figure 4: Check of thermodynamic scaling for the equation of state of o-terphenyl: (P/ρT+aρ/T) is plotted versus $\rho T^{-1/x}$ for several isotherms from 276K to 346K with 3 different pairs of adjustable parameters (*x, a*). The data collapse is acceptable is the 3 cases, but better in the right panel (*x*=6, *a*=800).

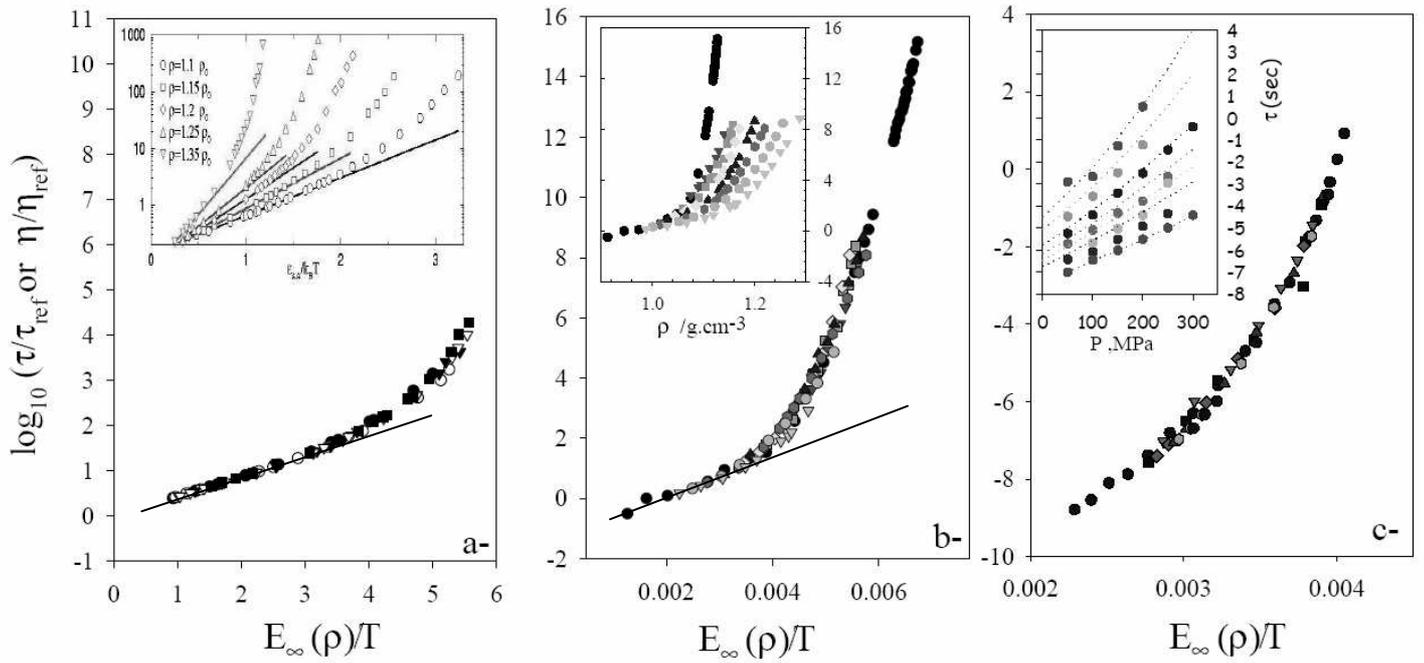

Fig.1

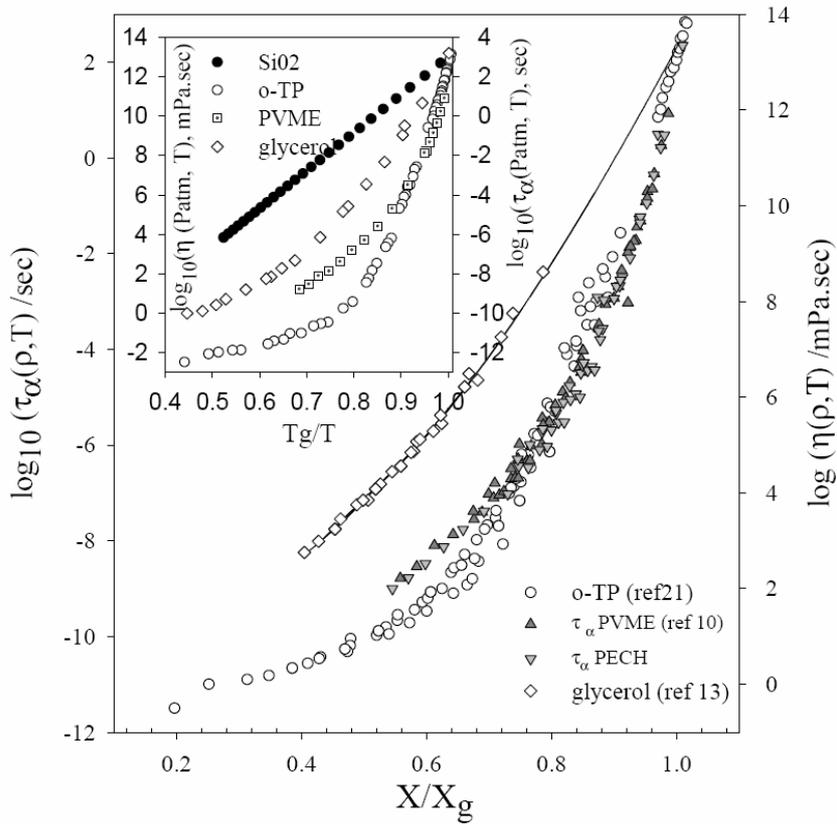

Fig. 2

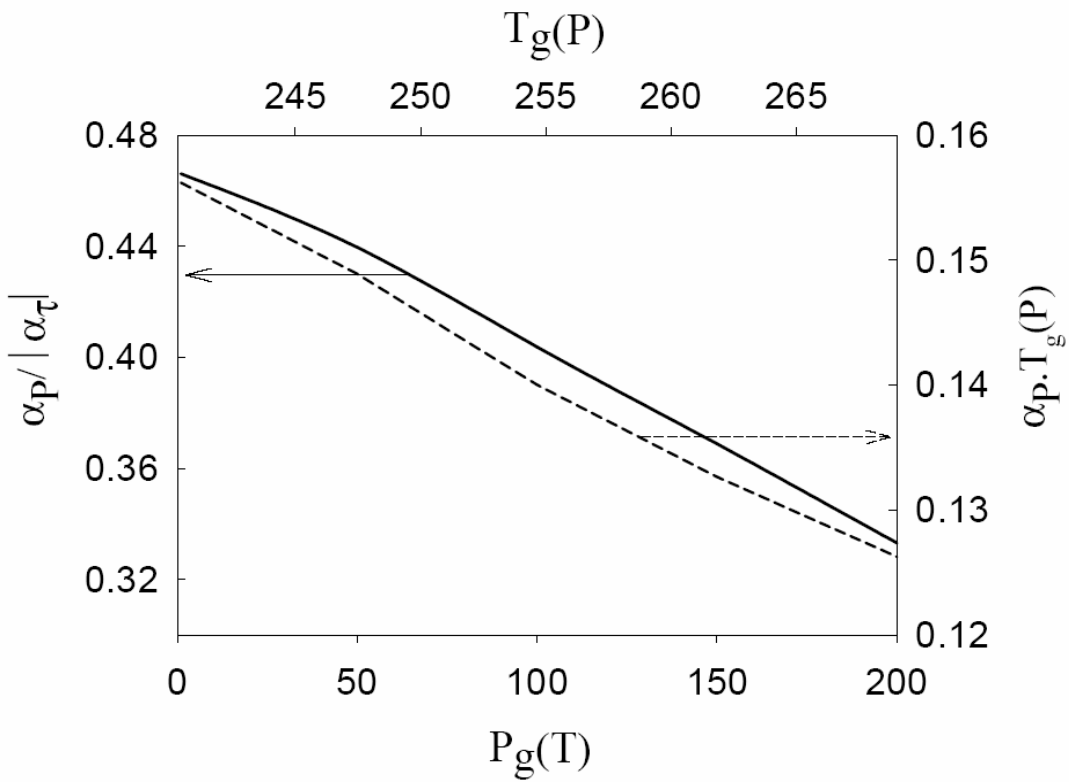

Fig.3

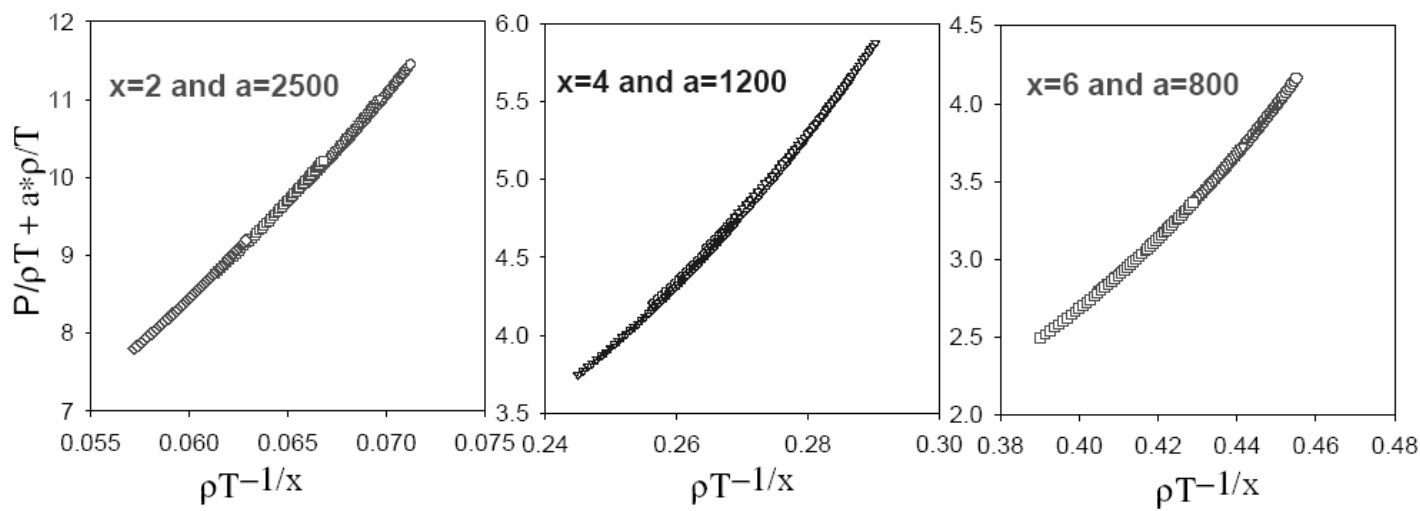

Fig.4